\documentclass[twocolumn,letter]{jpsj3}
\title{%
Dirac Electrons on a Sharply Edged Surface of Topological Insulators
}

\author{%
Yositake {\sc Takane} and Ken-Ichiro {\sc Imura}
}

\inst{%
Department of Quantum Matter, Graduate School of Advanced Sciences of Matter,\\
Hiroshima University, Higashihiroshima, Hiroshima 739-8530, Japan
}

\recdate{ \hspace{50mm} }

\abst{%
An unpaired gapless Dirac electron emergent at the surface of
a strong topological insulator (STI) is protected by
the bulk-surface correspondence
and believed to be immune to backward scattering.
It is less obvious, however, and yet to be verified explicitly whether
such a gapless Dirac state is smoothly extended over the entire surface
when the surface is composed of more than a single facet
with different orientations in contact with one another at sharp corner edges
(typically forming a steplike structure).
In the realistic situation that we consider, the anisotropy of the sample
leads to different group velocities in each of such facets.
Here, we propose that much insight on this issue can be obtained
by studying the electronic states on a hyperbolic surface of an STI.
By explicitly constructing the surface effective Hamiltonian,
we demonstrate that no backward scattering takes place
at a concave $90^\circ$ step edge.
A strong renormalization of the velocity
in the close vicinity of the step edge is also suggested.
}

\kword{%
strong topological insulator, Dirac electron, step edge, scattering problem
}

\begin{document}
\sloppy
\maketitle

A single Dirac cone emergent on the surface of
a three-dimensional (3D) strong topological insulator (STI) 
is ``topologically'' protected;
its existence guaranteed by the bulk-surface correspondence
and by a nontrivial value of the bulk topological invariant 
of the strong $\mathbb{Z}_2$ index.~\cite{fu,moore}
The existence of such a gapless Dirac cone has been repeatedly
verified even experimentally by 
a number of spin-resolved ARPES measurements performed
in different realizations of the 3D STI~\cite{ARPES0,ARPES1,ARPES2}
and is now incontrovertible.~\cite{hasan-kane}
Theorists have also predicted that, unlike weak topological insulators,
an STI exhibits a single surface Dirac cone
irrespective of the orientation of the surface.
However, the ARPES measurements can be carried out
only on the perfectly cleaved surface of layered STI samples,
i.e., ARPES is restricted to surfaces of some particular
easy-to-cleave orientations.
It is indeed a much less trivial issue whether the gapless Dirac
state observed on such one ``good'' facet of a crystal is smoothly extended to
adjacent ones, eventually covering the entire surface of the sample.
Experimentally, such a behavior of the surface Dirac state
may be accessible by STM measurements.
In ref.~\citen{yazdani}, an example of such an experiment
performed on atomic-scale terraces of an STI has been reported.

The protected gapless surface Dirac state is known to be robust against
perturbations that do not break time-reversal symmetry.
An electron in a protected surface state exhibits a notable feature, that is,
its spin is locked to a particular direction determined by its momentum
(spin-to-momentum locking).
This implies that
the spin direction of an electronic state with wave number $\mib{k}$
is orthogonal to that with wave number $-\mib{k}$, 
resulting in the complete suppression of $180^\circ$ backward scattering.
Here we concern ourselves with the transmission through
an interface of two Dirac electron systems,~\cite{takahashi}
in which the absence of $180^\circ$ backward scattering naturally
plays a central role.
Generally, gapless Dirac electrons, such as those found in graphene,
are not necessarily immune to backward scattering
at the interface~\cite{raoux,concha}
if the incoming electron is away from the normal incidence.
Only an electron normally incident to the interface
is forced to be completely transmitted
by the absence of $180^\circ$ backward scattering.
The greater the deviation from the normal incidence,
the larger the reflection becomes.

\begin{figure}[htp]
\begin{center}
\includegraphics[height=3.0cm]{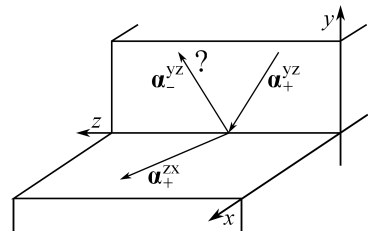}
\end{center}
\caption{Concave $90^\circ$ step edge consisting of two flat surfaces of
a 3D STI: one surface is on the $yz$ plane, and the other on the $zx$ plane,
and they meet at the $z$-axis.}
\end{figure}
Let us focus on the behavior of the protected surface state
on the L-shaped wall of an STI (see Fig. 1).
This structure can be regarded as a concave $90^\circ$ step edge
consisting of horizontal and vertical facets.
Note that both the horizontal terrace and the vertical
wall exhibit a protected gapless Dirac cone,
since they both separate an STI and a vacuum,
two topologically distinct worlds.
Although an STI exhibits a protected gapless Dirac cone
on surfaces of an arbitrary orientation, it can have, and in practice,
it always exhibits anisotropy in its model parameters.
As a natural consequence of this,
Dirac cones on facets of different orientations
(here, those on the vertical and horizontal surfaces)
generally have different apertures exhibiting different group velocities.
With this taken into account,
are these two Dirac cones still smoothly connected without being
affected by scattering at the interface,
even away from the normal incidence?
Naturally, this is an issue closely related to the corresponding problem
in the junction of two pure 2D Dirac systems such as in graphene.
In the case of the bulk-surface system we consider,
is there any difference
in the immunity of the Dirac electrons from scattering processes?
These are the issues we would like to address in this letter.
We set $\hbar = 1$ below.

To state the problem unambiguously,
let us first consider the surface electronic states 
in the asymptotic regions sufficiently away from the step edge.
We focus on a single concave $90^\circ$ step edge
consisting of two semi-infinite surfaces,
one on the $yz$ plane and the other on the $zx$ plane (see Fig.~1).
The system is assumed to be translationally invariant in the $z$-direction.
By treating each asymptotic region individually,
the effective Hamiltonians on the $yz$ and $zx$ planes can be
respectively deduced as
\begin{align}
 {\cal H}_{\rm eff}^{yz}
 &= \left[ \begin{array}{cc}
            0 & -A_{y}p_{y} + {\rm i}Bp_{z} \\
            -A_{y}p_{y} - {\rm i}Bp_{z} & 0
          \end{array}
   \right],
\\
 {\cal H}_{\rm eff}^{zx}
 &= \left[ \begin{array}{cc}
            0 & A_{x}p_{x} + {\rm i}Bp_{z} \\
            A_{x}p_{x} - {\rm i}Bp_{z} & 0
          \end{array}
   \right].
\end{align}
We take into account the anisotropy of the (bulk) system,
which results in the  velocity mismatch, i.e., if $A_{x} \neq A_{y}$.
We focus on eigenstates with energy $E = [(Ak)^{2}+(Bk_{z})]^{1/2}$,
where $Ak \equiv A_{y}k_{y} = A_{x}k_{x}$.
The corresponding eigenfunctions are
\begin{align}
     \label{eigenstate_1D_yz}
   {\mib\alpha}_{\pm}^{yz}
 & = \sqrt{\frac{1}{2A_{y}}}
     \left[
       \begin{array}{c}
         1 \\ \pm {\rm e}^{\mp{\rm i}\chi}
       \end{array}
     \right]
     {\rm e}^{\mp {\rm i}k_{y}y + {\rm i}k_{z}z } ,
              \\
     \label{eigenstate_1D_zx}
   {\mib\alpha}_{\pm}^{zx}
 & = \sqrt{\frac{1}{2A_{x}}}
     \left[
       \begin{array}{c}
         1 \\ \pm {\rm e}^{\mp{\rm i}\chi}
       \end{array}
     \right]
     {\rm e}^{\pm {\rm i}k_{x}x + {\rm i}k_{z}z } ,
\end{align}
where ${\rm e}^{\pm{\rm i}\chi} = (Ak \pm {\rm i}Bk_{z})/E$
and the prefactor is introduced to normalize the probability current
in the direction perpendicular to the $z$-axis.
Here, the sign $\pm$ indicates the direction of propagation (see Fig.~1).
Let us consider the scattering problem in which ${\mib\alpha}_{+}^{yz}$
is incident from $y = \infty$.
If $k_{z} = 0$, the reflection into ${\mib\alpha}_{-}^{yz}$ is forbidden
owing to the absence of $180^\circ$ backward scattering,
and hence the incident wave is
completely transmitted to ${\mib\alpha}_{+}^{zx}$.
Here comes the question to be answered:~\cite{comment_1}
does reflection occur when $k_{z} \neq 0$?

Note that solving a scattering problem at a junction of two inequivalent
(velocity-mismatched) Dirac systems is a formidable task.
Since one of our primary interests is to examine the continuity of
the current density at the corner, we need to deal with a subtle issue
related to the discontinuity of the wave functions, which is frequently
encountered in the study of Dirac systems.~\cite{takahashi,raoux,concha}
To overcome such a difficulty,
we introduce the following ``hyperbolic model''
(the details of the model specified later), in which
the surface of our 3D STI system with a $90^\circ$ concave step edge
is regarded as a limiting form of a hyperbola (Fig.~2).
The two asymptotes of the hyperbolic surface represent
the horizontal terrace and the vertical wall meeting at
the concave right angle of the STI crystal.
Using this hyperbolic model and
a set of continuous curvilinear coordinates in terms of this hyperbola,
we derive the surface effective Hamiltonian valid over the entire surface,
i.e., this provides us with a scheme for treating the two asymptotic surfaces
in the framework of a single Hamiltonian.~\cite{comment_2}
Note that in the conventional approach,
surface electronic states in the two asymptotic regimes,
such as those given by eqs.~(\ref{eigenstate_1D_yz}) and
(\ref{eigenstate_1D_zx}), are derived separately
as a solution of two individual uncorrelated problems.
Here, we treat them on the same footing
to discuss (eventually) how they are connected at the step edge.
The treatment of the opposite $90^\circ$ convex step edge
introduces further technical complications in the proposed approach.
This last issue will be briefly discussed toward the end of the paper.

Let us start with the following bulk effective Hamiltonian
for 3D anisotropic STIs in the continuum limit:~\cite{liu,shan}
$H_{\rm bulk} = m ({\mib p}) \tau_{z}
+ (A_{x}p_{x}\sigma_{x}+A_{y}p_{y}\sigma_{y}+Bp_{z}\sigma_z)\tau_{x}$,
where $p_{i}=-{\rm i}\partial_{i}$ ($i = x,y,z$) and
$m({\mib p}) = m_0 + m_2 (p_x^2 + p_y^2 + p_z^2)$ is the mass term.
Without loss of generality, we assume that $m_{0}>0$ and $m_{2}<0$.
The two types of Pauli matrices
$\mib{\sigma} =(\sigma_x, \sigma_y, \sigma_z)$ and
$\mib{\tau} =(\tau_x, \tau_y, \tau_z)$
respectively represent the real and orbital spin degrees of freedom.
If the ordinary matrix representation of $\mib{\sigma}$ is used,
$H_{\rm bulk}$ is expressed as
\begin{align}
       \label{matH}
   H_{\rm bulk}
   = \left[
   \begin{array}{cc}
             m ({\mib p})\tau_{z}+Bp_{z}\tau_{x} & {\mathcal D}_{-}\tau_{x} \\
             {\mathcal D}_{+}\tau_{x} & m ({\mib p})\tau_{z}-Bp_{z}\tau_{x}
           \end{array}
     \right],
\end{align}
where ${\mathcal D}_{\pm}=A_{x}p_{x}\pm{\rm i}A_{y} p_{y}$.
It should be noted that ${\cal H}_{\rm eff}^{yz}$ and
${\cal H}_{\rm eff}^{zx}$ can be derived from $H_{\rm bulk}$.~\cite{liu,shan}

\begin{figure}[btp]
\begin{center}
\includegraphics[height=5.0cm]{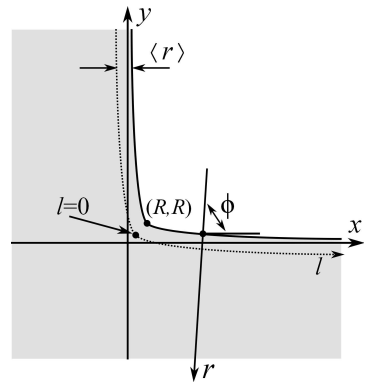}
\end{center}
\caption{Cross section (on the $xy$ plane) of our ``hyperbolic model'',
representing a curved surface of a 3D STI.
The cross section of the hyperbolic surface is given by $xy = R^{2}$,
reproducing a $90^\circ$ step edge
in the limit of sufficiently small $R$.}
\end{figure}
With the basic assumptions stated, let us introduce an ingenious trick.
As mentioned earlier, the main point of our protocol is to consider,
instead of directly treating the $90^\circ$ step edge,
the hyperbolic system depicted in Fig.~2.
We assume that the intersection of the curved surface with the $xy$ plane is
a rectangular hyperbola, $xy = R^{2}$,
and that the 3D STI is translationally invariant in the $z$-direction.
If $R$ is sufficiently small,
this surface can be regarded as a concave $90^\circ$ step edge.
It is convenient to introduce curvilinear coordinates, as shown in Fig.~2.
Let us draw a straight line that perpendicularly crosses the hyperbola
at the crossing point $(x_{0},y_{0})$.
We define $\phi$ as the angle between the $x$-axis and this line,
and $r$ as the distance from the crossing point.
The Cartesian coordinates $(x,y)$ are expressed as
\begin{align}
  x & = -r\cos\phi + x_{0}(\phi) ,
           \\
  y & = -r\sin\phi + y_{0}(\phi) ,
\end{align}
where $x_{0} \equiv R\tan^{\frac{1}{2}}\phi$
and $y_{0} \equiv R\cot^{\frac{1}{2}}\phi$.
The infinitesimal cross-sectional area is given by
${\rm d}S = (r + f(\phi)){\rm d}r{\rm d}\phi$, where
\begin{align}
       \label{eq:def-f}
  f = \sqrt{\left(\partial_{\phi} x_{0}\right)^{2}
            + \left(\partial_{\phi} y_{0}\right)^{2}}
    = \frac{R}{2\sin^{\frac{3}{2}}\phi \cos^{\frac{3}{2}}\phi} .
\end{align}
The derivatives are expressed as
\begin{align}
     \label{eq:partial_x}
  \partial_{x}
  & = - \cos\phi \partial_{r}
      + (r+f)^{-1}\sin\phi\partial_{\phi} ,
           \\
     \label{eq:partial_y}
  \partial_{y}
  & = - \sin\phi \partial_{r}
      - (r+f)^{-1}\cos\phi\partial_{\phi} .
\end{align}
With these expressions, we can rewrite the mass term as
$m ({\mib p}) = m_{\perp} + m_{\parallel}$ with
\begin{align}
  m_{\perp} &
  = m_{0} - m_{2}\left[ \partial_{r}^{2} + (r+f)^{-1}\partial_{r} \right] ,
               \\
  m_{\parallel} &
  = - m_{2}\left[ (r+f)^{-2}\partial_{\phi}^{2}
                  - (r+f)^{-3}(\partial_{\phi}f) \partial_{\phi}
           \right] .
\end{align}
Similarly, ${\mathcal D}_{\pm}$ is rewritten as
${\mathcal D}_{\pm} = \Omega_{\pm}
+ {\rm i}A_{\phi}{\rm e}^{\pm{\rm i}\tilde{\phi}}\partial_{r}$, where
\begin{align}
 & \Omega_{\pm}
   = {\rm i}(r+f)^{-1}
     \left(-A_{x}\sin\phi \pm {\rm i}A_{y}\cos\phi \right)\partial_{\phi} ,
          \\
 & {\rm e}^{\pm{\rm i}\tilde{\phi}}
   = \left(A_{x}\cos\phi \pm {\rm i}A_{y}\sin\phi\right)/A_{\phi}
\end{align}
with $A_{\phi}= (A_{x}^{2}\cos^{2}\phi+A_{y}^{2}\sin^{2}\phi)^{1/2}$.

To derive the surface effective Hamiltonian
in the spirit of $k\cdot p$-approximation,
we divide $H_{\rm bulk}$ into two parts as
$H_{\rm bulk} = H_\perp + H_{\parallel}$, where
\begin{align}
       \label{H_perp}
   H_{\perp}
 & = \left[
 \begin{array}{cc}
             m_{\perp}\tau_{z} & {\rm i}A_{\phi}
             {\rm e}^{-{\rm i}\tilde{\phi}}\partial_{r}\tau_{x} \\
             {\rm i}A_{\phi}{\rm e}^{{\rm i}\tilde{\phi}}\partial_{r}
             \tau_{x} & m_{\perp}\tau_{z}
           \end{array}
     \right] ,
           \\
       \label{delta_H}
   H_{\parallel}
 & = \left[
 \begin{array}{cc}
             m_{\parallel}\tau_{z}+Bp_{z}\tau_{x} & \Omega_{-}\tau_{x} \\
             \Omega_{+}\tau_{x} & m_{\parallel}\tau_{z}-Bp_{z}\tau_{x}
           \end{array}
     \right].
\end{align}
We first solve the radial eigenvalue equation~\cite{liu,shan,imura1,imura2}
$H_\perp |\psi \rangle = E_\perp |\psi \rangle$
with the boundary condition $|\psi (r=0)\rangle=\mib 0$,
that is, all four components of the wave function
$|\psi\rangle$ vanish on the surface.
As the simplest approximation,
we replace $(r+f)^{-1}\partial_{r}$ in $m_{\perp}$
with $\langle(r+f)^{-1}\rangle\partial_{r}$,
where the definition of the average $\langle(r+f)^{-1}\rangle$
is given below [see eq.~(\ref{eq:av_(r+f)_inv})].
Then, we can show that the eigenvalue equation
has surface solutions of the damped form,
$|\psi \rangle = {\rm e}^{-\kappa r} |u \rangle$,
where $\kappa^{-1}$ measures the penetration of
the surface wave function into the bulk.
By superposing two damped solutions,~\cite{comment_3}
we construct the solution of the radial eigenvalue equation
localized near the surface as
$|\Psi \rangle
={\rm e}^{-\kappa_{-}r}|u_{-}\rangle-{\rm e}^{-\kappa_{+}r}|u_{+}\rangle$.
The boundary condition $|\Psi (r=0)\rangle=\mib 0$ holds only when
$|u_{-}\rangle = |u_{+}\rangle$ for $\kappa_{+} \neq \kappa_{-}$. 
As shown in ref.~\citen{imura2}, this results in
the zero-energy condition $E_{\perp} = 0$ in our model.
We thus find that $\kappa_{\pm}$ is given by
\begin{align}
  \kappa_{\pm}(\phi)
  = \left({\cal A}_{\phi}
    \pm \sqrt{{\cal A}_{\phi}^{2}+4m_{0}m_{2}}\right)/(-2m_{2})
\end{align}
with ${\cal A}_{\phi} \equiv A_{\phi}-m_{2}\langle(r+f)^{-1}\rangle$.
We also find that two basis eigenstates, $|+ \rangle\rangle$
and $|- \rangle\rangle$, for $H_\perp$ with zero eigenvalue are given by
\begin{align}
   |\pm \rangle\rangle = \frac{1}{\sqrt{c_{\phi}}}
                         \rho (r,\phi) |\hat{\mib n} \pm \rangle ,
\end{align}
where $\rho (r,\phi) = {\rm e}^{-\kappa_{-}r} -{\rm e}^{-\kappa_{+}r}$,
$c_{\phi}$ is a $\phi$-dependent normalization constant, and
\begin{align}
        \label{dv_basis}
  |\hat{\mib n} \pm \rangle
   = \frac{1}{2}
     \left[
       \begin{array}{c}
         {\rm e}^{-{\rm i}\tilde{\phi} /2}
         \left(
           \begin{array}{c}
             1 \\ \mp{\rm i}
           \end{array}
         \right) \\
        \mp{\rm e}^{{\rm i}\tilde{\phi} /2}
         \left(
           \begin{array}{c}
             1 \\ \mp{\rm i}
           \end{array}
         \right)
       \end{array}
     \right] .
\end{align}
The real-spin sector $|S_{\pm}\rangle \equiv
\,^{t}({\rm e}^{-{\rm i}\tilde{\phi}/2},
\mp{\rm e}^{{\rm i}\tilde{\phi}/2})/\sqrt{2}$
points in the $\pm s_{x}$ direction in the regime of $y_{0} \to \infty$
($\phi \to 0$), while it points in the $\pm s_{y}$
direction in the regime of $x_{0} \to \infty$ ($\phi \to \pi/2$).
It should be emphasized that when $R$ is very small, $|S_{\pm}\rangle$
rapidly changes its direction from $\pm s_{x}$ to $\pm s_{y}$
across the point $(x_{0},y_{0})=(R,R)$.

Within the $k\cdot p$-approximation,
any surface state $|\mib\psi \rangle\rangle$ can be represented
as a linear combination of $|+ \rangle\rangle$ and $|- \rangle\rangle$
with the amplitude respectively specified by
$\alpha_+$ and $\alpha_-$, i.e.,
$|\mib\psi\rangle\rangle
= \alpha_+ |+ \rangle\rangle + \alpha_- |- \rangle\rangle$.
The effective surface Hamiltonian $\tilde{\cal H}_{\rm eff}$
for the two-component spinor
$\tilde{\mib\alpha} =\, ^{t}(\alpha_{+}, \alpha_{-})$ is defined by
\begin{align}
  \tilde{\cal H}_{\rm eff}
  = \left[
      \begin{array}{cc}
        \langle\langle +| H_{\parallel} |+ \rangle\rangle &
        \langle\langle +| H_{\parallel} |- \rangle\rangle \\
        \langle\langle -| H_{\parallel} |+ \rangle\rangle &
        \langle\langle -| H_{\parallel} |- \rangle\rangle
      \end{array}
    \right].
\end{align}
Here, each matrix element is expressed by
\begin{align}
  \langle\langle \sigma| H_{\parallel} |\sigma' \rangle\rangle
  = \int_{0}^{\infty}{\rm d}r \left(r+f\right)
    \frac{\rho(r,\phi)}{\sqrt{c_{\phi}}}
    \langle \hat{\mib n} \sigma | H_{\parallel} |\hat{\mib n} \sigma' \rangle
    \frac{\rho(r,\phi)}{\sqrt{c_{\phi}}} ,
\end{align}
where the factor $r+f$ reflects the fact
that ${\rm d}S = (r + f){\rm d}r{\rm d}\phi$
and $c_{\phi} = \int_{0}^{\infty}{\rm d}r (r+f)\rho(r,\phi)^{2}$.
We easily find that $\langle\langle \pm|H_{\parallel}|\pm \rangle\rangle = 0$.
In evaluating the off-diagonal elements,
we should note that $\partial_{\phi}$ in $H_{\parallel}$ acts on
not only $|\hat{\mib n} \sigma' \rangle$
but also $\rho(r,\phi)/\sqrt{c_{\phi}}$.
After tedious but straightforward calculations, we find
\begin{align}
     \label{h2}
  \tilde{\cal H}_{\rm eff}
  = \left[
      \begin{array}{cc}
        0 & \langle\langle +| H_{\parallel} |- \rangle\rangle \\
        \langle\langle -| H_{\parallel} |+ \rangle\rangle & 0
      \end{array}
    \right] ,
\end{align}
where
\begin{align}
     \label{eq:elements_off}
       \langle\langle \pm| H_{\parallel} |\mp \rangle\rangle
   = -{\rm i}\frac{\tilde A}{\langle r \rangle+f}\partial_{\phi}
     -\frac{{\rm i}}{2}\partial_{\phi}
      \left(\frac{\tilde A}{\langle r \rangle+f}\right) \pm {\rm i}Bp_{z}
\end{align}
with
\begin{align}
      \label{eq:ren-A}
   {\tilde A}(\phi)
   = \left[A_{\phi}-m_{2}\langle (r+f)^{-1} \rangle\right]
     \partial_{\phi} \tilde{\phi}.
\end{align}
In these equations, we have used the notation
\begin{align}
        \label{eq:av_r}
 & \langle r \rangle
   = \frac{\int_{0}^{\infty}{\rm d}r r \rho(r,\phi)^{2}}
          {\int_{0}^{\infty}{\rm d}r \rho(r,\phi)^{2}} ,
       \\
        \label{eq:av_(r+f)_inv}
 & \langle (r+f)^{-1} \rangle
   = \frac{\int_{0}^{\infty}{\rm d}r (r+f)^{-1} \rho(r,\phi)^{2}}
          {\int_{0}^{\infty}{\rm d}r \rho(r,\phi)^{2}} .
\end{align}
The second term on the right-hand side of eq.~(\ref{eq:elements_off}) is
essential in ensuring the hermiticity of $\tilde{\cal H}_{\rm eff}$.

It is convenient to introduce the one-dimensional coordinate
\begin{align}
      \label{eq:1D-coordinate}
  l = \int_{\pi/4}^{\phi}{\rm d}\phi'
      \left[\langle r \rangle (\phi') + f(\phi') \right] ,
\end{align}
located slightly beneath the (geometrical) surface (see Fig.~2).
Note that the limit of $\phi \to 0$ ($\pi/2$) corresponds to
$l \to -\infty$ ($+\infty$).
We rewrite the effective Hamiltonian with this coordinate.
Since ${\rm d}l = (\langle r \rangle +f){\rm d}\phi$,
the wave function $\mib\alpha(l)$ in the new coordinate is related to
the old one by
$\mib\alpha(l) \equiv \tilde{\mib\alpha}(\phi)/(\langle r \rangle +f)^{1/2}$.
From eq.~(\ref{h2}), we find that the effective Hamiltonian
for $\mib\alpha(l)$ is expressed as
\begin{align}
&      \label{h-1D}
  {\cal H}_{\rm eff} = 
  \nonumber \\
&  \left[
      \begin{array}{cc}
        0 & {\tilde A}p_{l} + \frac{1}{2} \left(p_{l}{\tilde A}\right)
            +{\rm i}Bp_{z} \\
        {\tilde A}p_{l} + \frac{1}{2} \left(p_{l}{\tilde A}\right)
        -{\rm i}Bp_{z} & 0
      \end{array}
    \right]
\end{align}
with $p_{l} \equiv -{\rm i}\partial_{l}$.
This form of an effective Dirac Hamiltonian
has been suggested in ref.~\citen{takahashi}.
The probability current operator in the $l$ direction is given by
\begin{align}
      \label{current-1D}
  j_{l}
  = \left[
      \begin{array}{cc}
        0 & {\tilde A} \\
        {\tilde A} & 0
      \end{array}
    \right].
\end{align}
Let us consider the behavior of ${\tilde A}(l)$ given in eq.~(\ref{eq:ren-A}),
which should be regarded as the effective velocity
along the one-dimensional coordinate.
Since $f \to \infty$ in the limit of $l \to \pm\infty$
[see eq.~(\ref{eq:def-f})
and the note given below eq.~(\ref{eq:1D-coordinate})],
we observe with the aid of
$\partial_{\phi} \tilde{\phi} = A_{x}A_{y}/A_{\phi}^{2}$
that ${\tilde A}(- \infty) = A_{y}$ and ${\tilde A}(\infty) = A_{x}$.
Contrastingly, the velocity can become very large in the close vicinity of
the step edge if $R$ is vanishingly small.
Note that $m_{2} < 0$ is assumed from the outset,
and that $f \approx 0$ near $l = 0$ (i.e., $\phi = \pi/4$)
if $R$ is very small.
The correction term $-m_{2}\langle (r+f)^{-1} \rangle$
in ${\tilde A}$ becomes very large,
leading to a large increase in velocity.

Now, we can answer the main question raised earlier.
The solution of ${\cal H}_{\rm eff}\mib\alpha = E\mib\alpha$
with $E = [(Ak)^{2}+(Bk_{z})]^{1/2}$ is obtained as
\begin{align}
     \label{eigenstate_1D}
  {\mib\alpha}_{\pm}
  = \sqrt{\frac{1}{2{\tilde A}(l)}}
    \left[
      \begin{array}{c}
        1 \\ \pm {\rm e}^{\mp{\rm i}\chi}
      \end{array}
    \right]
    \exp\left( \pm {\rm i} \int_{0}^{l}{\rm d}l'\frac{Ak}{{\tilde A}(l')}
               + {\rm i}k_{z}z \right) .
\end{align}
We see that eq.~(\ref{eigenstate_1D}) respectively reproduces
eqs.~(\ref{eigenstate_1D_yz}) and (\ref{eigenstate_1D_zx})
in the regimes of $l \ll -R$ and $l \gg R$.
We also see from eq.~(\ref{current-1D}) that the prefactor
${\tilde A}(l)^{-1/2}$ guarantees current conservation over the entire system.
That is, this ${\mib\alpha}_{\pm}$ continuously connects
the two asymptotic eigenfunctions.~\cite{comment_4}
In addition, eq.~(\ref{eigenstate_1D}) is justified
even for vanishingly small (but finite) $R$.
Therefore, we conclude that
no reflection takes place at a concave $90^\circ$ step edge,
although the velocity becomes very large in its vicinity
and hence the amplitude of the wave function is reduced.

Let us now consider an inverted situation in which
the convex side of the hyperbola (see Fig.~2) is filled with a 3D STI.
In this situation, our curvilinear coordinates
($r < 0$ and $0 < \phi < \pi/2$) can be applied only to a limited region
near the hyperbola of width much smaller than $R$.
Indeed, $\partial_{x}$ and $\partial_{y}$ in eqs.~(\ref{eq:partial_x})
and (\ref{eq:partial_y}) become ill-defined
owing to the presence of $(-|r|+f)^{-1}$ if $|r| \gtrsim R$.
Therefore, we restrict our analysis on surface states to the case
where $R$ is sufficiently larger than
the penetration depth $\langle |r| \rangle$.
Analysis similar to that reported above reveals that surface states
in the convex case are described by an effective Hamiltonian
essentially equivalent to eq.~(\ref{h-1D}) but the velocity, now given by
${\tilde A}(\phi)=\left[A_{\phi}+m_{2}\langle (-|r|+f)^{-1} \rangle\right]
\partial_{\phi} \tilde{\phi}$,
is reduced [note the sign change in front of the correction term
compared with that in eq.~(\ref{eq:ren-A})].
This implies that the behavior of surface states in the convex case
is essentially equivalent to that in the concave case
except that the velocity is renormalized in the opposite way.

We finally comment on the validity of our analysis.
Since $H_{\rm bulk}$ is valid in the long-wavelength regime,
one may question whether it can be applied to
STIs with a sharply edged surface.
In eq.~(\ref{eigenstate_1D}),
the shortest length scale of the derived eigenfunction
is on the order of $\langle r \rangle$
even though $R$ becomes vanishingly small.
Thus, we expect that our approach will be justified as long as
$\langle r \rangle$ is much longer than the lattice constant.
This does not necessarily mean that the employed approach allows
quantitative predictions of the electronic properties
of, for example, atomic-scale terraces of an STI with atomic-scale precision.
Yet, our conclusion (no reflection at a single step edge)
itself is consistent with the experimental result indicating
that topological surface states are transmitted
through atomic-scale steps with high probabilities.~\cite{yazdani}

\section*{Acknowledgment}

{\small
The authors are supported by KAKENHI:
Y.T. by a Grant-in-Aid for Scientific Research (C) (No. 24540375)
and K.I. by the ``Topological Quantum Phenomena'' (No. 23103511).}


\begin{thebibliography}{99}
\bibitem{comment_1}
The authors of ref.~\citen{moon} have analyzed this type of problem
by requiring the continuity of wave functions at the interface.
However, this procedure is questionable
since different spin-quantization axes are implicitly used in expressing
${\mib\alpha}_{\pm}^{yz}$ and ${\mib\alpha}_{\pm}^{zx}$.
Indeed, the spin of ${\mib\alpha}_{\pm}^{yz}$ (${\mib\alpha}_{\pm}^{zx}$)
is perpendicular to the $s_{x}$-axis ($s_{y}$-axis) in the spin space.
\bibitem{comment_3}
The radial eigenvalue equation gives
four exponentially decreasing (and increasing) solutions. 
We use the former to construct the two eigenstates of $H_{\perp}$ 
compatible with the boundary condition.~\cite{liu,shan,imura1,imura2}
\bibitem{comment_4}
This result itself is in a sense consistent with the analysis of
ref.~\citen{moon} (see also the comments in ref.~\citen{comment_1}).
\end{thebibliography}
\end{document}